# Transport evidence of the three-dimensional Dirac semimetal phase in doped α-Sn grown by molecular beam epitaxy

Yuanfeng Ding,[1] Bingxin Li,[1] Chen Li,[1] Yan-Bin Chen,[2] Hong Lu[1,3,4,†] and Yan-Feng Chen[1]

[1]*National Laboratory of Solid State Microstructures & Department of Materials Science and Engineering, College of Engineering and Applied Sciences, Nanjing University, Nanjing 210093, China*

[2]*Department of Physics, Nanjing University, Nanjing 210093, China*

[3]*Jiangsu Key Laboratory of Artificial Functional Materials, Nanjing University, Nanjing 210093, China*

[4]*Collaborative Innovation Center of Advanced Microstructures, Nanjing University, Nanjing 210093, China*

[†]Corresponding author: hlu@nju.edu.cn

# ABSTRACT

We report the quantum transport properties of the $\alpha$-Sn films grown on CdTe (001) substrates by molecular beam epitaxy. The $\alpha$-Sn films are doped with phosphorus to tune the Fermi level and access the bulk state. Clear Shubnikov-de Haas oscillations can be observed below 30 K and a nontrivial Berry phase has been confirmed. A nearly spherical Fermi surface has been demonstrated by angle-dependent oscillation frequencies. In addition, the sign of negative magnetoresistance which is attributed to the chiral anomaly has also been observed. These results provide strong evidence of the three-dimensional Dirac semimetal phase in $\alpha$-Sn.

Three-dimensional (3D) Dirac semimetal can be viewed as a 3D analogue to graphene which holds Dirac fermions in bulk. Many materials, such as $\beta$-BiO$_2$ [1], Bi$_{1-x}$Sb$_x$ [2], $A_3$Bi ($A$ = Na, K, Pb), [3] and Cd$_3$As$_2$ [4] have been theoretically predicted to be 3D Dirac semimetals. The most famous examples are Cd$_3$As$_2$ and Na$_3$Bi. The Dirac linear energy dispersion with Dirac points in the bulk Brillouin zone has been confirmed by angle-resolved photoemission spectroscopy (ARPES) [5-8]. The transport measurement is another important method to detect the topological properties of topological semimetals. Large linear magnetoresistance [9,10], negative magnetoresistance [11-14], 3D quantum Hall effect [15-17], and pressure-induced superconductivity [18,19] have been observed in these materials. In particular, a nontrivial Berry phase, which can be extracted from Shubnikov-de Haas (SdH) oscillations, is a distinguished feature of Dirac fermions [14,20-22]. A critical issue in the studies of Dirac semimetals is to explore new material systems. At present most Dirac semimetal materials are suffered from complicated crystal structures and constituents or instability in the ambient atmosphere. For further investigations on Dirac semimetals, high-quality and stable samples are necessary.

Gray tin, also known as $\alpha$-Sn, is predicted to be a Dirac semimetal or topological insulator with strain [23-25]. In the past decade, the topological band structures of $\alpha$-Sn have been progressively studied [24-33]. As an elemental topological material, $\alpha$-Sn has a diamond crystal structure and simple constituent and is thermally stable when epitaxially grown on InSb or CdTe substrates [34,35]. The epitaxial $\alpha$-Sn films on these substrates are compressively strained and should be a Dirac semimetal according to predictions [24,25]. The 3D Dirac band dispersions have been observed by ARPES in 30-bilayer *α*-Sn (111) [31]. Several transport studies also support the nontrivial topology of *α*-Sn [36-39]. Thicker films are preferable to suppress quantum confinement effect, therefore the 3D Dirac semimetal phase can be more clearly demonstrated by a sphere Fermi surface in bulk [38]. However, direct evidence of 3D Dirac cone *α*-Sn(001) measured by ARPES is still lacking due to the small momentum separation in Γ-Z [40]. Besides, other important transport properties, such as chiral-anomaly-induced negative magnetoresistance (NMR) and planar Hall effect (PHE), have never been reported in *α*-Sn. This is probably due to the shunting effect of InSb substrates and limited film quality. The shunting effect can be solved using CdTe instead of InSb as substrates. The spin-polarized topological surface state of *α*-Sn films on CdTe substrates has been confirmed by transport method [37]. Due to the relatively low Fermi level, the information about the bulk Dirac points has not been obtained. Therefore, it is interesting to explore the Dirac semimetal phase in *α*-Sn by tuning the Fermi level.

In this letter, we have grown a series of *α*-Sn films doped by phosphorus on CdTe (001) substrates by molecular beam epitaxy (MBE). The quantum transport properties are systematically studied on the *α*-Sn sample whose Fermi level is close to the bulk Dirac points. A nontrivial Berry phase and a nearly spherical Fermi surface are revealed, providing strong evidence for the 3D Dirac semimetal phase in *α*-Sn. In addition, the sign of NMR and PHE which is attributed to the chiral anomaly is observed. These results indicate *α*-Sn is a promising Dirac semimetal for further investigations.

The doped α-Sn samples were grown by an MBE system (Dr. Eberl MBE-Komponenten Octoplus 300) with a high-purity Sn source (99.9999%). Before the growth of α-Sn, an InSb buffer needs to be grown on CdTe (001) substrate in a III-V MBE system (Veeco GENxplor) to improve the surface quality. The method of substrate treatment and buffer growth have been discussed detailedly in our previous work [37]. The growth rate of α-Sn is about 0.1 Å/s and the growth temperature is decreased well below room temperature by a cooling plate. The α-Sn layer is 50 nm thick and doped by phosphorus in-situ to tune the Fermi level.. In Figure 1 (a) and (b), a clear and streaky (2×2) surface reconstruction, which indicates the formation of α-Sn, can be observed. The X-Ray diffraction (XRD) pattern in Figure 1 (c) shows a clear α-Sn (004) peak without any signal of β-Sn. Figure 1 (d) is the reciprocal space mapping (RSM) on the (115) plane of the α-Sn sample, which indicates that the α-Sn is fully strained to the CdTe substrate. The in-plane constant of the α-Sn sample is the same as the CdTe constant, while the out-plane constant is about 6.499 Å. We can deduce the uniaxial tensile strain of ~ 0.15% along the [001] axis of the α-Sn film, which should be a Dirac semimetal as predicted [24].

The electrical measurement was performed in a physical property measurement system (PPMS, Quantum Design) with the lowest temperature of 2 K and magnetic field up to 9 T. The α-Sn is 5×5 mm$^2$ square with a four-probe configuration and Ag paste contacts. Magneto-transport measurements were performed on α-Sn with different doping concentrations[]. In the following paragraph, we will focus on the α-Sn with a nominal doping concentration of $1\times10^{20}$ cm$^{-3}$.

Magneto-transport property of doped α-Sn is shown in Figure 2. The magnetic field is perpendicular to the sample surface and the results are acquired by averaging the data measured at positive and negative fields. Figure 2 (a) shows magnetoresistance (MR) curves of the sample at different temperatures and the inset of Figure 2 (a) is the temperature-dependent resistance of the sample at zero field. The MR is parabolic at low field and becomes linear as the field increases, which indicates a linear energy dispersion of Dirac fermions. Figure 2 (b) shows the Hall resistance of the sample. The nearly linear field dependence of Hall resistance and negative Hall coefficient means that the transport is dominated by electrons. Besides, we can observe the SdH oscillation in both MR and Hall resistance at 2 K even though the background has not been subtracted. These features of MR and Hall resistance indicate that the Fermi level is lifted to the conduction band by P doping and close enough to the Dirac point. To investigate the concentration and mobility of the carrier, we fitted the Hall data by the two-band model:

$$\rho_{yx} = \frac{B}{e} \frac{(n_1\mu_1^2 + n_2\mu_2^2) + \mu_1^2\mu_2^2 B^2 (n_1 + n_2)}{(n_1\mu_1 + n_2\mu_2)^2 + \mu_1^2\mu_2^2 B^2 (n_1 + n_2)^2} \quad (1)$$

where $\rho_{yx} = R_{yx}/d$ is the Hall resistivity, $d$ is the thickness of the α-Sn film, $n_1, \mu_1, n_2, \mu_2$ represent the carrier density and mobility of two kinds of carriers, respectively, e is the elementary charge. The fitting results are shown in Figure 2 (c) and (d), $n_2$ increased rapidly when temperature higher than 100 K with a relatively low mobility $\mu_2$ (~200 cm$^2$ V$^{-1}$ s$^{-1}$). It corresponds to the decreasing of longitudinal resistance when the temperature is higher than 150 K as shown in the inset of Figure 2 (a). This phenomenon is consistent with the intra-valley and inter-valley excitation of the trivial carriers in α-Sn [41]. The $n_1$ with higher mobility $\mu_1$ suggests another electron channel which will be discussed later.

To identify the topological properties of the α-Sn film, we investigate its SdH oscillation by subtracting the polynomial background from the MR curve. Figure 3 (a) shows the SdH oscillation subtracted background at 2 K. According to the suggestion of Ando [42], we used the oscillation of magneto-conductance to extract the Berry phase and it can be described by the Lifshitz-Kosevich formula [43]:

$$\Delta G_{xx} \propto \cos 2\pi \left(\frac{F}{B} - \gamma + \delta\right) \tag{2}$$

Where $\gamma = \frac{1}{2} - \frac{\phi_B}{2\pi}$ is related to Berry phase $\phi_B$, the phase factor $\delta$ is 0 for 2D Fermi surface while ±1/8 for 3D Fermi surface [44] and F is the oscillation frequency. The oscillation minimum (maximum) corresponds to the integer (half-integer) Landau index N. The Landau index is also plotted in Figure 3 (a) and linearly fitted by:

$$2\pi \left(\frac{F}{B} - \gamma + \delta\right) = (2N - 1)\pi \tag{3}$$

The fitting result shows an intercept of 0.6 which reveals a nontrivial Berry Phase of -0.8π. It means that SdH oscillation in α-Sn originates from Dirac fermions. Then we investigated the shape of the Fermi surface by angle-dependent magneto-transport measurement. As shown in Figure 3(b), θ is defined as the angle between the magnetic field and sample surface normal. Figure 3(c) shows the angle-dependent MR of the doped α-Sn sample at 2 K. The MR decreases monotonically as the θ increases. When θ approaches 90°, we can find in the inset of Figure 3(c) that the MR decreases with increasing magnetic field when *B*>4 T, which is the sign of NMR induced by chiral anomaly. We extracted the SdH oscillation from MR at different θ and obtained the oscillation frequency by fast Fourier transform (FFT). The results are shown in Figure 3(d) and the inset. The oscillation frequency almost maintains near 25 T at different $\theta$, indicating a 3D spherical Fermi surface. It can be found in the inset of Figure 3 (c) that the oscillation frequency increases slightly when $\theta$ is about 50°. Such anomaly may due to the anisotropy of the $\Gamma_8^+$ band in $\alpha$-Sn. The Fermi vector along [111] direction ($\theta\sim54.7°$) is longer than others. We can derive the carrier concentration by SdH oscillation frequency $F$ using the following equation:

$$n_{SdH} = \frac{1}{3\pi^2}\left(\frac{2eF}{\hbar}\right)^{\frac{3}{2}} \quad (4)$$

the $n_{SdH}$ is $7.10\times10^{17}$ cm$^{-3}$. This value is consistent with $n_1$ at 2 K ($7.68\times10^{17}$ cm$^{-3}$) as shown in Figure 2(c), which means that $n_1$ in Figure 2(c) is attributed to Dirac fermions from 3D Dirac cone. Besides, the high mobility $\mu_1$ ($\sim2192$ cm$^2$ V$^{-1}$ s$^{-1}$ at 2 K) also indicates that $n_1$ should be Dirac fermions. The nontrivial Berry phase and 3D spherical Fermi surface are strong transport evidence of the Dirac semimetal phase in $\alpha$-Sn. It also indicates that the Fermi level has been tuned close to the bulk Dirac point. Here we can have an estimate of the Fermi level. In a Dirac cone, Fermi-level can be expressed as $E_F = \hbar k_F v_F$, where the Fermi vector $k_F = (3\pi^2 n_{SdH})^{1/3}$. Assuming a Fermi velocity of $5\times10^5$ m/s [45], we can deduce that the Fermi level is about 91 meV above bulk Dirac point.

As another important transport property of Dirac semimetal, NMR induced by chiral anomaly has never been reported in $\alpha$-Sn before. In order to determine the origin of NMR in the doped $\alpha$-Sn shown in Figure 3(c), we measured the MR with the magnetic field rotated in the plane of the sample. The angle $\varphi$ is the angle between the magnetic field and current as defined in Figure 4(a). In Figure 4(b), we can find that the NMR is sensitive to the angle $\varphi$. As the magnetic field rotating in the plane of the sample, NMR vanished rapidly when $\varphi$ is above 30°. This is because the positive MR becomes stronger as the component of magnetic field perpendicular to current increasing. Such a feature is consistent with NMR induced by chiral anomaly in other topological semimetals such as TaAs and $Cd_3As_2$ [12,13,46,47]. The temperature-dependent NMR is shown in Figure 4 (c). The NMR can be observed when $B>4$ T and T⩽100 K and vanished at high temperatures. To further analysis quantitively, the NMR is fitted by the semiclassical formula [11,48,49]:

$$\sigma = (1 + C_w B^2)\sigma_{WAL} + \sigma_n \tag{5}$$

$$\sigma_{WAL} = a\sqrt{B} + \sigma_0 \tag{6}$$

where $\sigma_{WAL}$ is the conductivity related to weak antilocalization, $\sigma_0$ is the conductivity under zero field, $\sigma_n$ corresponds to the contribution of the conventional nonlinear band near the Fermi level. $C_w B^2$ is the term related to the chiral anomaly, where $C_w$ is a positive coefficient. The fitting result is shown in the inset of Figure 4 (b). $C_w$ at 2 K is 5.77×10$^{-4}$ T$^{-2}$ which is much smaller than the value in other topological semimetals [11,46,50]. Usually, the positive MR at low field (B<1 T) is caused by weak antilocalization (WAL) when the magnetic field and current are parallel, However, we note that the positive MR in our sample is quite different because the WAL will disappear as the increasing magnetic field will break the time-reversal symmetry. Such a strong positive MR may be attributed to the electronics from the trivial band because the T-dependence of NMR in Figure 4(c) is quite similar to the thermal excitation behavior of $n_2$ fitted by Hall data. Although there is no Lorentz force in parallel field theoretically, the electric field and magnetic may be aligned imperfectly in experiment and cause a finite positive MR. Therefore the value of $C_w$ is quite smaller than others. However, we think that such a fitting can show the signature of NMR induced by chiral anomaly to some degree. We can also exclude other origins of NMR in of sample. Firstly, NMR can be found in magnetic materials [51]. However, $α$-Sn is a non-magnetic material. Besides, the current jetting effect [52] due to the geometry of the contacts can also be neglected. Lastly, NMR can also arise in a trivial band in ultraquantum limit [53] ($ωτ$>>1, $ω$ is the cyclotron frequency and $τ$ is the transport relaxation time). In our sample, $ωτ$~0.8 at 4 T and 1.8 at 9 T which indicates that the sample is not under ultraquantum limit.

Another signature of chiral anomaly which is called the planar Hall effect has been predicted and observed in many topological semimetals in recent years [54-57]. To provide more evidence of chiral anomaly in $α$-Sn, we measured the planar Hall effect (PHE) and the anisotropic magneto-resistance (AMR) of the sample as shown in Figure 4 (d) and (e). All the data were measured under both positive and negative magnetic fields to eliminate the normal Hall effect induced by misalignment. The AMR and PHE can be well-fitted by the following equations [54,57]:

$$R_{yx}^{PHE} = -\Delta R \sin\theta \cos\theta \qquad (7)$$

$$R_{xx}^{AMR} = R_\perp - \Delta R \cos^2\theta \qquad (8)$$

where $\Delta R = R_\perp - R_\parallel$, $R_\perp$ and $R_\parallel$ correspond to the longitudinal resistance when the magnetic field is perpendicular and parallel to the current direction, respectively. Due to the little misalignment of the contacts, the PHE has a resistivity deviation from zero. So the eq. (7) can be corrected as [56]:

$$R_{yx}^{PHE} = -\Delta R \sin\theta \cos\theta + a(R_\perp - \Delta R\cos^2\theta) \qquad (9)$$

In the similar way, the eq. (8) can be corrected as:

$$R_{xx}^{AMR} = R_\perp - \Delta R\cos^2\theta - b\Delta R \sin\theta \cos\theta \qquad (10)$$

The solid curves in Figure 4 (d) and (e) are the fitting result. The fitted $\Delta R$ depended on the magnetic field is shown in Figure 4(f). $\Delta R$ with a positive amplitude increases monotonically as the magnetic field increases. The PHE and AMR have also been found in ferromagnetic materials [58] and topological insulators with spin-polarized topological surface states [37,59] with a negative amplitude ($\Delta R < 0$). Our $\alpha$-Sn sample has no ferromagnetic order and positive amplitude ($\Delta R > 0$). Therefore, the PHE and AMR don't originate from topological surface states or ferromagnetism. In topological semimetals, the PHE and AMR has been confirmed originating from chiral anomaly with positive amplitude. It is consistent with our result in Figure 4(f). Therefore, we believe that the AMR and PHE in our sample are attribute to chiral anomaly. Although, an asymmetric Fermi-surface of real metal may result in a positive orbital MR [60], we suppose that the Fermi-surface in our sample is spherical as shown in Figure 3 (c). The observation of both NMR and PHE shows strong evidence of chiral anomaly in $\alpha$-Sn.

In conclusion, we have grown a series $\alpha$-Sn films by MBE and tuned the Fermi level to access the bulk Dirac point by phosphorus doping. Magneto-transport measurement results show a nontrivial Berry phase of -0.8$\pi$ and nearly spherical Fermi surface. The NMR and PHE with positive amplitude was observed in $\alpha$-Sn for the first time, which are the signature of chiral anomaly. All of these transport evidence support that *α*-Sn is a Dirac semimetal for future application.

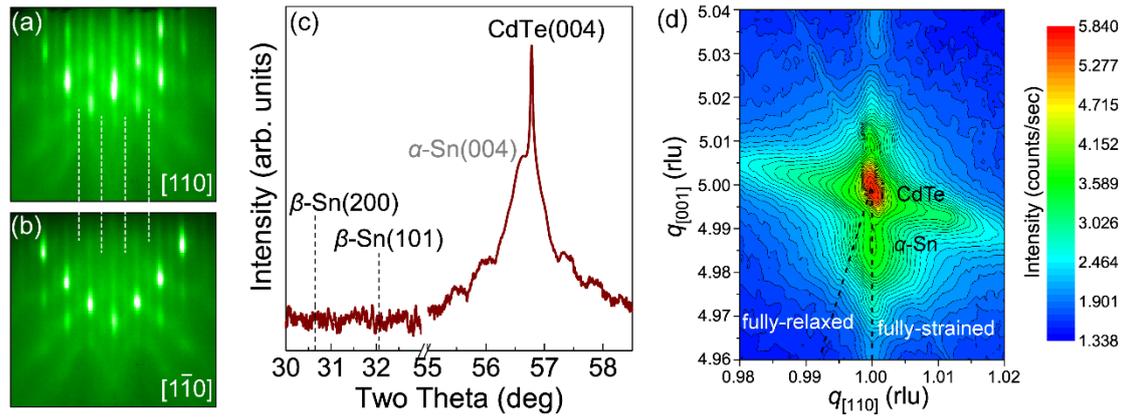

FIG 1 Structural characterization of the α-Sn sample. (a) and (b) RHEED patterns of the α-Sn sample. White dashed line shows the (2×2) reconstruction of α-Sn. (c) X-ray diffraction pattern of the α-Sn sample. (d) Reciprocal space mapping on the (115) plane of the α-Sn sample.

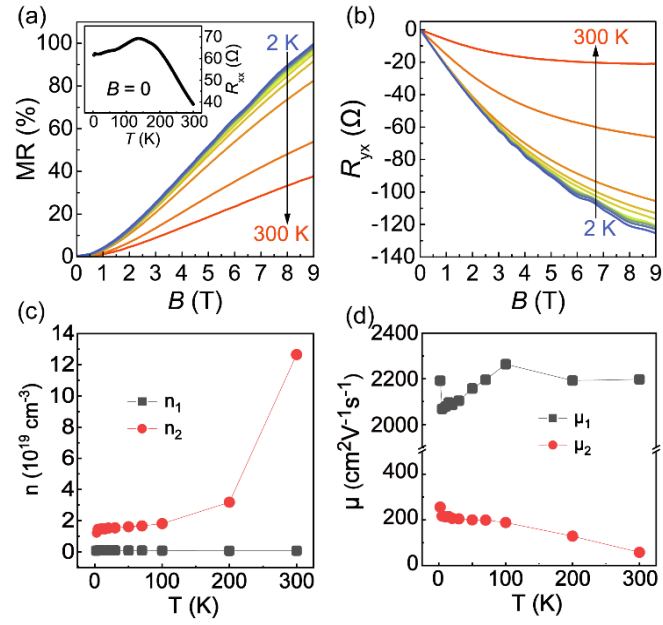

FIG 2 Magneto-transport properties of doped α-Sn. (a) The magneto-resistance curves at different temperatures. (b) The Hall resistance curves at different temperatures. (c) and (d) The carrier densities and mobilities fitted by two-band model, respectively.

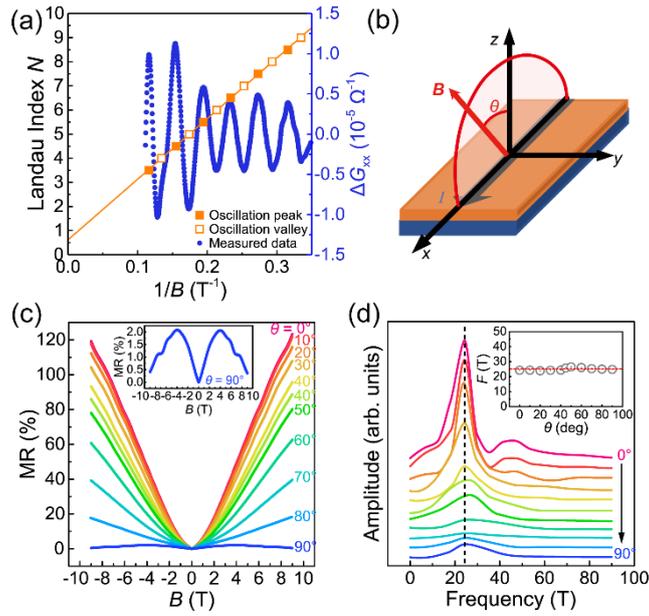

FIG. 3 Angle-dependent magneto-transport properties of the α-Sn sample. (a) the SdH oscillation and Landau index at 2 K with magnetic field perpendicular to sample surface. (b) The schematic diagram of angle-dependent measurement with the magnetic fields rotating in x-z plane. x is defined as the direction of current and z is the normal direction of sample surface. $\theta$ is the angle between the magnetic field and z axis. (c) The magneto-resistance curves at different $\theta$. The inset shows the magneto-resistance when magnetic field is parallel to the current. (d) FFT result of SdH oscillation at different $\theta$ subtracted from magneto-resistance in (c). The inset show the frequency of SdH oscillation at different $\theta$.

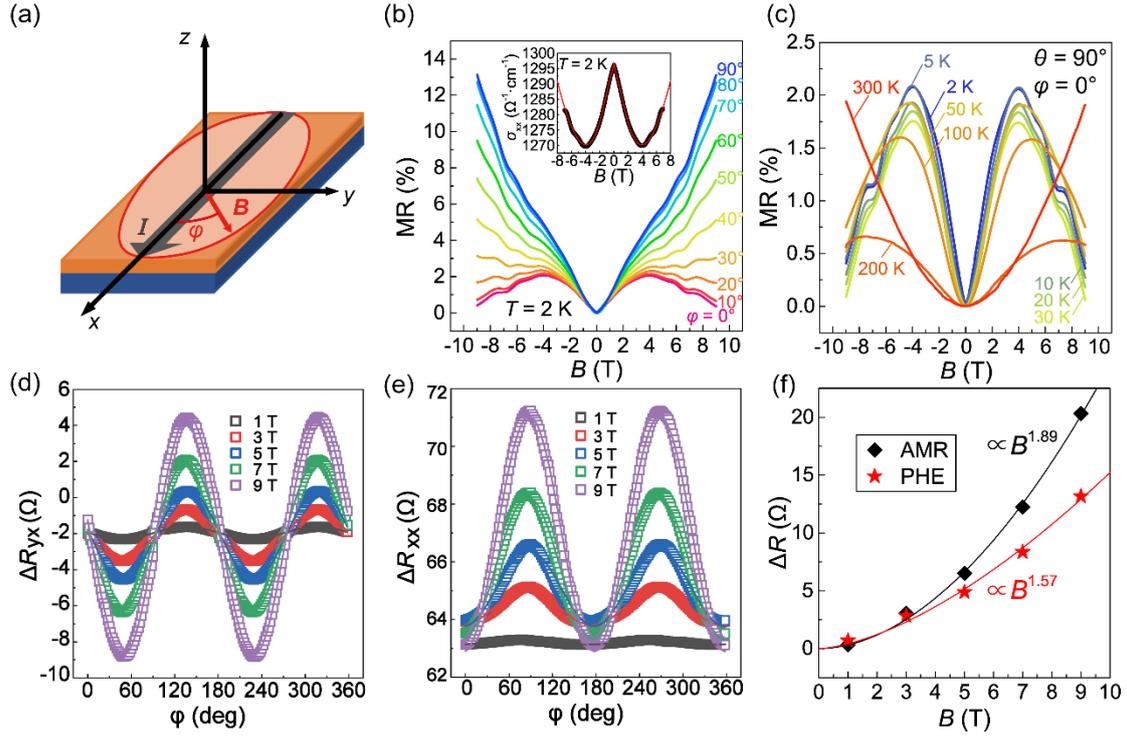

FIG. 4 Angle-dependent magneto-transport measurement with the magnetic field rotating in the sample surface ($\theta=90°$). (a) The schematic diagram of angle-dependent measurement with the magnetic fields rotating in the sample surface. x is defined as the direction of current. $\varphi$ is defined as the angle between the magnetic field and the current direction. (b) $\varphi$-dependent magneto-resistance curves at 2 K. The inset shows the magneto-conductance at $\varphi=0°$. Red line is the fitting results by eq. (5) and (6). (c) Temperature-dependent magneto-resistance curves with the magnetic field parallel to the current direction. (d) and (e) $\varphi$-dependent transverse resistance (PHE) and longitudinal resistance (AMR), respectively. The solid curves are the fitting results using eq. (9) and (10), respectively. (f) Magnetic field-dependent amplitude of (PHE) and (AMR) acquired from the fitting results in (d) and (e).